\documentclass[onecolumn,usenatbib,useAMS]{mn2e}

\usepackage{natbib}
\usepackage{graphicx}
\usepackage{latexsym,amssymb}
\usepackage[fleqn]{amsmath}

\newcommand\avg[1]{\left\langle{#1}\right\rangle}

\newcommand\like{{\cal L}}
\newcommand\Lbark{\bar\like^{(k)}}
\newcommand\vol{{\cal V}}
\newcommand\Nnest{N_{\rm nest}}
\newcommand\Ntot{N_{\rm tot}}
\newcommand\Zlive{Z^{\rm live}}
\newcommand\Ztot{Z^{\rm tot}}

\newcommand\refeq[1]{eq.~(\ref{eq:#1})}
\newcommand\refeqs[2]{eqs.~(\ref{eq:#1}) and (\ref{eq:#2})}
\newcommand\reffig[1]{Figure~\ref{fig:#1}}

\newcommand\refsec[1]{\S~\ref{S:#1}}
\newcommand\refsecs[2]{\S\S~\ref{S:#1} and \ref{S:#2}}

\title[Statistical uncertainty in nested sampling]
{On statistical uncertainty in nested sampling}

\author[Keeton]{
Charles R.\ Keeton \\
Department of Physics and Astronomy, Rutgers University, 136 Frelinghuysen Road, Piscataway, NJ 08854 USA
}

\begin{document}


\maketitle 

\begin{abstract}
Nested sampling has emerged as a valuable tool for Bayesian analysis, in particular for determining the Bayesian evidence.  The method is based on a specific type of random sampling of the likelihood function and prior volume of the parameter space.  I study the statistical uncertainty in the evidence computed with nested sampling.  I examine the uncertainty estimator from \citet{Skilling04,Skilling06} and introduce a new estimator based on a detailed analysis of the statistical properties of nested sampling.  Both perform well in test cases and make it possible to obtain the statistical uncertainty in the evidence with no additional computational cost.
\end{abstract}


\section{Introduction}
\label{S:intro}

Bayesian statistics provide a general framework for confronting models with data \citep[e.g.,][]{Gelman}.  Constraints on model parameters are quantified by the \emph{posterior distribution} for the parameters given the data.  The overall quality of a model is characterised by an integral over the posterior, which is known as the \emph{evidence}.  The Bayesian evidence is especially valuable as an objective means of comparing models with different numbers of parameters.

The challenge with Bayesian statistics is finding an efficient method to explore the posterior and/or compute the evidence.  The posterior may occupy many dimensions and have a complicated (and possibly multi-modal) shape.  Markov Chain Monte Carlo (MCMC) methods have become popular as a way to generate samples of points drawn from arbitrary posteriors \citep[e.g.,][]{Gelman}.  MCMC samples are great for inferring parameter values and ranges, but they cannot be used by themselves to evaluate the evidence.  MCMC methods can be extended to yield the evidence via thermodynamic integration \citep[see][and references therein]{GelmanMeng}, but that approach can be computationally intensive.

\citet{Skilling04,Skilling06} recently introduced an approach called nested sampling that is specifically designed to compute the Bayesian evidence.  Roughly speaking, the idea is to peel away layers of constant likelihood one by one, and combine the likelihood values with the volumes of the layers to obtain the evidence.  The volumes may be difficult to determine, but they can be estimated statistically if the likelihood layers are chosen in a particular way (see \refsec{nestsamp} for details).  While the analysis focuses on the evidence, it can yield a set of points drawn from the posterior as a natural by-product.

There are  two practical challenges with nested sampling.  The first is that at each step we need to generate a new point drawn from the region inside an iso-likelihood surface.  A lot of the literature on nested sampling addresses methods for picking new points.  \citet{Mukherjee} discuss drawing points inside a multi-dimensional ellipsoid that encloses the likelihood surface at $\like_0$, and ignoring any that have $\like < \like_0$.  \citet{Shaw}, \citet{Feroz1}, and \citet{Feroz2} develop methods that use multiple ellipsoids to handle more complicated likelihood functions, including multi-modal distributions.  \citet{Chopin} point out that importance sampling can be powerful if one can find a distribution that is easy to draw from and approximates the likelihood distribution moderately well.  \citet{Betancourt} advocates using constrained Hamiltonian Monte Carlo methods to evolve a new point from one of the known points.  All of those methods keep the core approach of peeling away likelihood layers in sequence from the outside in, and differ only in the details of picking new points.  \citet{Brewer} introduce a variant they call diffusive nested sampling that does not always require the steps to proceed from the outside in.

The second challenge is that nested sampling, like any stochastic sampling procedure, has some statistical uncertainty in its results.  General proofs establish that nested sampling converges to the correct evidence with an error that scales as $N^{-1/2}$ where $N$ is a measure of the computational effort \citep{Chopin,Skilling09}.  However, in practical applications it would be nice to have a specific estimate of the statistical uncertainty in the evidence.  That is the purpose of this paper.  I first review the nested sampling procedure (\refsec{nestsamp}) and a popular estimator from \citet{Skilling04,Skilling06} for the statistical uncertainty in the evidence (\refsec{skilling}).  I then introduce a new uncertainty estimator based on an analysis of the statistical properties of the nested sampling procedure (\refsecs{moments}{Zlive}).  I use numerical tests to assess the estimators and provide some guidelines for choosing parameters that control the number of samplings (\refsec{num}).  The results presented here are applicable to any implementation of nested sampling that uses the conventional approach of peeling away likelihood layers in one direction only (i.e., to all current methods other than diffusive nested sampling).

\section{Theoretical Framework}
\label{S:framework}

\subsection{Nested sampling}
\label{S:nestsamp}

To establish the concepts and notation, it is useful to review the nested sampling algorithm (see \citealt{Skilling04,Skilling06} for details).  Consider a likelihood function $\like(\theta)$ defined on a parameter space $\theta$, which may be multi-dimensional.\footnote{To simplify the notation, I do not explicitly indicate vectors or write the data dependence in the likelihood function.}  Priors on the parameters are specified by $\pi(\theta)$, which is normalised such that $\int \pi(\theta)\ d\theta = 1$.  With simple flat priors, $\pi(\theta) = 1/V$ where $V$ is the volume spanned by the allowed range of parameters, but the framework can incorporate non-flat priors as well.  The Bayesian evidence is then
\begin{equation} \label{eq:evid1}
  Z = \int \like(\theta)\ \pi(\theta)\ d\theta
\end{equation}
Define a function $X(L)$ to be the fraction of the prior volume that lies at a likelihood level higher than $L$:
\begin{equation}
  X(L) = \int_{\like(\theta)>L} \pi(\theta)\ d\theta
\end{equation}
This is a monotonic decreasing function, with $X(0)=1$.  In principle, we can invert to find $L(X)$ and then rewrite \refeq{evid1} as
\begin{equation} \label{eq:Xint}
  Z = \int_{0}^{1} L(X)\ dX
\end{equation}
Now suppose we can generate a sample of $\Nnest$ points $\{L_i,X_i\}$ such that the likelihood increases while the fractional volume decreases with the index $i$; in other words, $L_i > L_{i-1}$ and $X_i < X_{i-1}$, and we can consider $L_0=0$ and $X_0=1$.  Then we can discretise the integral to estimate the evidence as
\begin{equation} \label{eq:evidsum}
  Z = \sum_{i=1}^{\Nnest} L_i (X_{i-1}-X_i)
\end{equation}
Later it will be useful to consider the buildup of evidence by examining the ``partial evidence'' due to the contribution from the first $k$ steps:
\begin{equation} \label{eq:partevid}
  Z_k = \sum_{i=1}^{k} L_i (X_{i-1}-X_i)
\end{equation}
There is some error in \refeq{evidsum} associated with approximating the integral as a sum, but it is generally small compared with the statistical uncertainty \citep{Skilling06}.  There is also some error induced by truncating the sum, which is discussed in \refsec{Zlive}.

The heart of nested sampling is the method for generating the likelihood sampling $\{L_i\}$ and volume sampling $\{X_i\}$.  The idea is that it is (relatively) straightforward to produce a relevant likelihood sampling, but it can be difficult to determine the associated volumes so we treat those statistically.  Consider some likelihood threshold $\like_0$ enclosing a volume $\vol_0$.  Suppose we have $M$ points drawn \emph{uniformly} from that volume.  In general there will be some (slightly) higher likelihood threshold $\like_1 > \like_0$ that encloses all $M$ points.  Statistically speaking, we can estimate the smaller enclosed volume as $\vol_1 = \vol_0 t_1$ where $t_1$ is the largest of $M$ random numbers drawn uniformly between 0 and 1.  In other words, $t_1$ is drawn from the probability distribution for the largest of $M$ uniform deviates between 0 and 1, which is
\begin{equation} \label{eq:pt}
  p(t) = M\,t^{M-1} \quad\mbox{for}\ t \in [0,1]
\end{equation}
We can generalise to non-uniform priors by defining the ``volumes'' to be integrals of the priors over the relevant regions and having the $M$ points drawn from the prior distribution.  The probability distribution for $t_1$ remains unchanged.

That idea leads to the following procedure.  Begin with $M$ points---known as ``live'' points---drawn uniformly from the full prior distribution.  Let the likelihoods of the live points be $\like_\mu$ for $\mu=1,\ldots,M$.  Then at step $k$ of the nested sampling:
\begin{enumerate}
\item Extract the lowest likelihood live point and call it the $k$-th sampled point: $L_k = \min(\like_\mu)$.
\item Estimate the associated volume as
\begin{equation} \label{eq:Xk}
  X_k = X_{k-1} t_k
\end{equation}
where $t_k$ is a random number drawn from $p(t)$ in \refeq{pt}.
\item Replace the extracted live point with a new point that is drawn from the priors but restricted to the region $\like(\theta) \ge L_k$.
\end{enumerate}
Iterating this process for a total of $\Nnest$ steps yields a likelihood sampling $\{L_i\}$ and volume sampling $\{X_i\}$ that can be combined using \refeq{evidsum} to estimate the evidence.  This is the conventional nested sampling technique as defined by \citet{Skilling04,Skilling06}.  The various implementations of nested sampling mainly differ in the way they find the replacement point in step (iii).

\subsection{Skilling's error analysis}
\label{S:skilling}

To estimate the statistical uncertainty associated with stochastic sampling,  \citet{Skilling04,Skilling06} invokes information theory.  In general the posterior $p(\theta)$ is (much) narrower than the prior $\pi(\theta)$, and we can characterise the difference in terms of the ``information gain'' (also known as the Kullback-Leibler divergence; see \citealt{Kullback})
\begin{equation}
  H = \int p(\theta) \ln\frac{p(\theta)}{\pi(\theta)}\ d\theta
\end{equation}
By Bayes's theorem, $p(\theta) = \like(\theta)\pi(\theta)/Z$ so we can write
\begin{equation}
  H = \frac{1}{Z} \int \like(\theta) \pi(\theta) \ln\frac{\like(\theta)}{Z}\ d\theta
  = \frac{1}{Z} \int L(X) \ln L(X)\ dX \ -\ \ln Z
\end{equation}
using the same change of variables as in \refeq{Xint}.  This integral can be discretised just like the evidence integral, so it is straightforward to estimate $H$ from a given sampling $\{L_i,X_i\}$.

\citet{Skilling04,Skilling06} argues that the number of steps needed to reach the posterior is approximately $H M$ where $M$ is the number of live points, and that the dominant statistical uncertainty arises from Poisson fluctuations $\sqrt{H M}$ in the number of steps.  Thus, he estimates an uncertainty in $\ln Z$ of about $\sqrt{H/M}$.  Note that Skilling argues that $\ln Z$, and not $Z$ itself, is the quantity likely to have a fairly symmetric and quasi-Gaussian distribution.  However, if the uncertainty is small (specifically, $\sigma_Z/Z \ll 1$), then $Z$ itself will also be Gaussian distributed and Skilling's estimate corresponds to a fractional uncertainty in the evidence of
\begin{equation} \label{eq:Ssig1}
  \frac{\sigma_Z}{Z} \approx \sqrt{\frac{H}{M}}
\end{equation}
This estimator is often used in nested sampling applications, but its accuracy has not (to my knowledge) been rigorously established.

\subsection{Moment-based error analysis}
\label{S:moments}

\citet{Skilling06} mentions that it should be possible to obtain a more detailed estimate of the statistical uncertainty by computing the mean and variance of $Z$ over all possible realisations of the volume sampling $\{X_i\}$, but he does not carry out the analysis.  The goal of this section is to compute $\avg{Z}$ and $\avg{Z^2}$ to obtain a new estimator for $\sigma_Z$.  Since this estimator is based on the standard deviation, it is most useful when $Z$ is Gaussian distributed, i.e., when the uncertainties are small ($\sigma_Z/Z \ll 1$).  This does not seem like a significant limitation, though, because in many applications it will be desirable to achieve small uncertainties.

It is convenient to use \refeq{Xk} to write the volumes as
\begin{equation} \label{eq:Xprod}
  X_i = \prod_{j=1}^{i} t_j
\end{equation}
The advantage is that the $X_i$'s are statistically correlated, but the $t_i$'s are independent and that allows us to decompose the joint probability density for all the $t_i$'s into a product:
\begin{equation}
  p_{\rm all}(t_1,t_2,t_3,\ldots) = p(t_1)\ p(t_2)\ p(t_3) \cdots
\end{equation}
where $p(t)$ is from \refeq{pt}.  We can then write the average of any quantity $f$ over all realisations of the volume sampling as
\begin{equation}
  \avg{f} \equiv \int f(t_1,t_2,t_3,\ldots)\ p(t_1)\ p(t_2)\ p(t_3) \cdots\ dt_1\ dt_2\ dt_3 \cdots
\end{equation}
It is important to understand that such an average only spans the volume sampling; at this point we are not considering different realisations of the likelihood sampling.  As part of this analysis we need moments of the $t$ probability distribution,
\begin{equation}
  \avg{t^n} = \int_{0}^{1} t^n\ p(t)\ dt = \frac{M}{M+n}
\end{equation}

Combining \refeqs{evidsum}{Xprod}, we can write the (partial) evidence in terms of the $t_i$'s as
\begin{equation}
  Z_k = \sum_{i=1}^{k} L_i (1-t_i) \prod_{j=1}^{i-1} t_j
  = \sum_{i=1}^{k} L_i \left( \prod_{j=1}^{i-1} t_j - \prod_{j=1}^{i} t_j \right)
\end{equation}
Since the terms in the products are statistically independent, we can factorise the average of a product and write
\begin{equation}
  \avg{ \prod_{j=1}^{n} t_j } = \prod_{j=1}^{n} \avg{t_j} = \avg{t}^n
\end{equation}
for any $n \in [1,\Nnest]$.  This allows us to write the average of the evidence after any step $k$ as
\begin{equation}
  \avg{Z_k} = \avg{ \sum_{i=1}^{k} L_i \left( \prod_{j=1}^{i-1} t_j - \prod_{j=1}^{i} t_j \right) }
  = \sum_{i=1}^{k} L_i \left( \avg{t}^{i-1} - \avg{t}^i \right)
  = \frac{1}{M} \sum_{i=1}^{k} L_i \avg{t}^{i}
  = \frac{1}{M} \sum_{i=1}^{k} L_i \left(\frac{M}{M+1}\right)^{i}
  \label{eq:Zavg}
\end{equation}
This is a simple expression for the (partial) evidence averaged over all possible realisations of the volume sampling (given a particular likelihood sampling $\{L_i\}$).  Obviously the final evidence is obtained just by evaluating at $k=\Nnest$.

To compute the second moment it is convenient to begin with the partial evidence from \refeq{partevid}:
\begin{eqnarray}
  \avg{Z_k^2} &=& \avg{ \left[ \sum_{i=1}^{k} L_{i} (1-t_{i}) \prod_{j=1}^{i-1} t_{j} \right]
    \left[ \sum_{i'=1}^{k} L_{i'} (1-t_{i'}) \prod_{j'=1}^{i'-1} t_{j'} \right] } \nonumber\\
  &=& \avg{Z_{k-1}^2} + \avg{ \left[ L_k (1-t_k) \prod_{j=1}^{k-1} t_j \right]^2 }
    + 2 \avg{ \left[ \sum_{i=1}^{k-1} L_{i} (1-t_{i}) \prod_{j=1}^{i-1} t_{j} \right]
    \left[ L_k (1-t_k) \prod_{j'=1}^{k-1} t_{j'}\right] }
\label{eq:Zk2a}
\end{eqnarray}
In the second line I separate the joint sum over $i,i' \le k$ into three components.  The first component includes all terms with $i,i' \le k-1$, so we can immediately recognise it as $\avg{Z_{k-1}^2}$.  The second component is the term with $i=i'=k$.  The third component includes all terms in which one index equals $k$ while the other runs over values $\le k-1$.  Since we can interchange $i$ and $i'$, there is a leading factor of 2.

It takes a few steps to evaluate the averages.  First consider the second term in \refeq{Zk2a}.  Writing out the products of $t_i$'s and collecting terms yields
\begin{equation}
  \avg{ L_k^2 \left( \prod_{j=1}^{k-1} t_j^2 - 2 t_k \prod_{j=1}^{k-1} t_j^2
    + \prod_{j=1}^{k} t_j^2 \right) }
  = L_k^2 \left( \avg{t^2}^{k-1} - 2 \avg{t} \avg{t^2}^{k-1} + \avg{t^2}^k \right)
  = \frac{2}{M(M+1)} L_k^2 \avg{t^2}^{k}
  \label{eq:Zk2b}
\end{equation}
Now consider the third term in \refeq{Zk2a}.  We can rewrite the products, taking care to distinguish the $t$'s that appear twice in a product from those that appear just once, and thus obtain
\begin{eqnarray}
  &&\avg{ 2 L_k \sum_{i=1}^{k-1} L_i \left(
    \prod_{j=1}^{i-1} t_{j}^2 \prod_{j'=i}^{k-1} t_{j'}
  - \prod_{j=1}^{i} t_{j}^2 \prod_{j'=i+1}^{k-1} t_{j'}
  - \prod_{j=1}^{i-1} t_{j}^2 \prod_{j'=i}^{k} t_{j'}
  + \prod_{j=1}^{i} t_{j}^2 \prod_{j'=i+1}^{k} t_{j'}
  \right) } \nonumber\\
  &&\quad = 2 L_k \sum_{i=1}^{k-1} L_i \left(
    \avg{t^2}^{i-1} \avg{t} ^{k-i}
  - \avg{t^2}^{i} \avg{t}^{k-i-1}
  - \avg{t^2}^{i-1} \avg{t}^{k-i+1}
  + \avg{t^2}^{i} \avg{t}^{k-i}
  \right) \nonumber\\
  &&\quad = \frac{2}{M(M+1)} L_k \avg{t}^{k} \sum_{i=1}^{k-1} L_i \frac{\avg{t^2}^{i}}{\avg{t}^{i}}
  \label{eq:Zk2c}
\end{eqnarray}
Notice that \refeq{Zk2b} has the same form as each term in the sum in \refeq{Zk2c}, but with index $i=k$.  So when we insert \refeqs{Zk2b}{Zk2c} back into \refeq{Zk2a}, we can write
\begin{equation}
  \avg{Z_k^2} = \avg{Z_{k-1}^2}
    + \frac{2}{M(M+1)} L_k \avg{t}^{k} \sum_{i=1}^{k} L_i \frac{\avg{t^2}^{i}}{\avg{t}^{i}}
  \label{eq:Zk2avg}
\end{equation}
With this expression for the second moment of the partial evidence, we see that the second moment of the full evidence can be written as
\begin{equation}
  \avg{Z^2} = \frac{2}{M(M+1)} \sum_{k=1}^{\Nnest} L_k \avg{t}^{k}
    \sum_{i=1}^{k} L_i \frac{\avg{t^2}^{i}}{\avg{t}^{i}}
  = \frac{2}{M(M+1)} \sum_{k=1}^{\Nnest} L_k \left(\frac{M}{M+1}\right)^{k}
    \sum_{i=1}^{k} L_i \left(\frac{M+1}{M+2}\right)^{i}
  \label{eq:Z2avg}
\end{equation}
Combining \refeqs{Zavg}{Z2avg} in the usual way yields a new estimator for the statistical uncertainty in the evidence:
\begin{equation} \label{eq:Ksig}
  \sigma_Z^2 = \frac{2}{M(M+1)} \sum_{k=1}^{\Nnest} L_k \left(\frac{M}{M+1}\right)^{k}
    \sum_{i=1}^{k} L_i \left(\frac{M+1}{M+2}\right)^{i}
  - \frac{1}{M^2} \left[\sum_{i=1}^{\Nnest} L_i \left(\frac{M}{M+1}\right)^{i}\right]^2
\end{equation}
For comparison, rewriting \refeq{Ssig1} in the current notation yields the following expression for Skilling's uncertainty estimator:
\begin{equation} \label{eq:Ssig2}
  \sigma_Z^2 = \frac{1}{M^3} \left[\sum_{i=1}^{\Nnest} L_i \left(\frac{M}{M+1}\right)^{i}\right]
    \left[\sum_{j=1}^{\Nnest} L_j \ln L_j \left(\frac{M}{M+1}\right)^{j}\right]
  - \frac{1}{M^3} \left[\sum_{i=1}^{\Nnest} L_i \left(\frac{M}{M+1}\right)^{i}\right]^2
    \ln\left[\frac{1}{M} \sum_{j=1}^{\Nnest} L_j \left(\frac{M}{M+1}\right)^{j}\right]
\end{equation}
On the surface these two expressions look quite different, so it is interesting to compare them in quantitative examples.

\subsection{Handling the remainder}
\label{S:Zlive}

When the nested sampling procedure is complete, there is some (small) remaining volume, $X_{\Nnest}$, whose contribution to the evidence is neglected in \refeq{evidsum}.  While we can make its contribution arbitrarily small by taking enough steps (see \refsec{num-steps}), we can also include it at the expense of making the analytic expressions slightly more complicated.

Suppose we truncate nested sampling after step $k$ and compute the partial ``nested evidence'' $Z_k$ from \refeq{partevid}.  We can estimate the remaining evidence as a product of the remaining volume, $X_k$, and the mean likelihood within that volume.  Since the live points are drawn uniformly from $X_k$, we can estimate the mean likelihood from the live points as
\begin{equation}
  \Lbark = \frac{1}{M} \sum_{\mu=1}^{M} \like_\mu
\end{equation}
Here the overbar distinguishes this average over live points from an average over volume realisations, and the superscript is a reminder that the average is taken after step $k$.  Thus the ``live evidence'' is
\begin{equation} \label{eq:Zlive}
  \Zlive_k = \Lbark X_k
\end{equation}
Averaging the live evidence over all volume realisations yields
\begin{equation} \label{eq:avgZlive}
  \avg{\Zlive_k} = \Lbark \avg{X_k} = \Lbark \avg{\prod_{j=1}^{k} t_j} = \Lbark \left(\frac{M}{M+1}\right)^k
\end{equation}
The second moment is
\begin{equation}
  \avg{(\Zlive_k)^2} = (\Lbark)^2 \avg{X_k^2} = (\Lbark)^2 \avg{\prod_{j=1}^{k} t_j^2}
  = (\Lbark)^2 \left(\frac{M}{M+2}\right)^k
\end{equation}
so the statistical uncertainty in the live evidence is
\begin{equation} \label{eq:sigZlive}
  \sigma_{\Zlive_k}^2 = (\Lbark)^2 \left(\frac{M}{M+1}\right)^k 
    \left[ \left(\frac{M+1}{M+2}\right)^k - \left(\frac{M}{M+1}\right)^k \right]
\end{equation}
Now consider the estimate of the total evidence after step $k$,
\begin{equation} \label{eq:avgZtot}
  \Ztot_k = Z_k + \Zlive_k
\end{equation}
The average over volume realisations is simply obtained from \refeqs{Zavg}{avgZlive}.  The statistical uncertainty in $\Ztot_k$ is
\begin{equation}
  \sigma_{\Ztot_k}^2 = \sigma_{Z_k}^2 + \sigma_{\Zlive_k}^2
    + 2 \left( \avg{Z_k \Zlive_k} - \avg{Z_k} \avg{\Zlive_k} \right)
\end{equation}
The term in parentheses accounts for the fact that $\Zlive_k$ and $Z_k$ are not independent because they both involve the same volume sampling.  The cross term can be evaluated using an analysis similar to that in \refeq{Zk2c}, which yields
\begin{equation}
  \avg{Z_k \Zlive_k} = \frac{\Lbark}{M+1} \left(\frac{M}{M+1}\right)^k
    \sum_{i=1}^{k} L_i \left(\frac{M+1}{M+2}\right)^i
\end{equation}
Putting the pieces together, we find that including the live evidence increases the statistical uncertainty in the total evidence according to
\begin{equation} \label{eq:sigZtot}
  \sigma_{\Ztot_k}^2 = \sigma_{Z_k}^2 + \Delta\sigma_{Z_k}^2
\end{equation}
where the original uncertainty is given in \refeq{Ksig} while the increase is
\begin{equation}
  \Delta\sigma_{Z_k}^2 = (\Lbark)^2 \left(\frac{M}{M+1}\right)^k
    \left[ \left(\frac{M+1}{M+2}\right)^k - \left(\frac{M}{M+1}\right)^k \right]
  + 2 \Lbark \left(\frac{M}{M+1}\right)^k \sum_{i=1}^{k} L_i
      \left[ \frac{1}{M+1}\left(\frac{M+1}{M+2}\right)^i - \frac{1}{M}\left(\frac{M}{M+1}\right)^i \right]
\end{equation}
In the examples that follow, I take enough steps that the live evidence provides a negligible contribution by the end, but the formalism in this section can be used if the number of nested sampling steps is more modest.

\section{Numerical Results}
\label{S:num}

In this section I present numerical tests designed to assess the uncertainty estimators, and to investigate how many samples to use.  Since the nested sampling framework does not require any specific assumptions about the form of the likelihood distribution, a Gaussian test case should be sufficient.  However, I also consider a log-normal distribution as a check.

\subsection{Gaussian test case}
\label{S:gaus}

Consider a multivariate Gaussian likelihood specified by some mean vector and covariance matrix.  With flat priors we can make the following simplifications.  Choose coordinates centred on the mean and aligned with the principal axes of the covariance matrix.  Scale each coordinate by the standard deviation in that direction.  This yields a multivariate Gaussian in canonical form,
\begin{equation}
  \like(\theta) = (2\pi)^{-d/2}\ e^{-|\theta|^2/2}
\end{equation}
where $d$ is the number of dimensions.  Let the prior volume be a cube of side length $s$ centred on the origin, so the prior volume is $V = s^d$ and the priors are $\pi(\theta) = 1/V$.  Thus the evidence is
\begin{equation}
  Z = V^{-1} \int_V (2\pi)^{-d/2}\ e^{-|\theta|^2/2}\ d\theta
\end{equation}
If the prior box is large enough to encompass essentially all of the likelihood, then $V Z \approx 1$ independent of the box size.  For this reason, in the following tests I examine $V Z$ instead of just $Z$.  The information gain for this case is
\begin{equation}
  H = \frac{1}{V Z} \int \like \ln\frac{\like}{Z}\ d\theta
  \approx -\frac{d}{2}\left(1+\ln 2\pi\right) - \ln Z
\end{equation}
In the last step I again assume the prior box is large.

For concreteness, I use a box with side length $s = 10$ in $d = 4$ dimensions; these choices influence the quantitative details but do not affect the general conclusions.  In the fiducial case I use $M = 400$ live points and take $\Nnest = 4100$ steps (see \refsec{num-steps}).  The associated information gain is $H=3.53$, and Skilling's estimator of the fractional uncertainty in the evidence has an analytic value of $\sqrt{H/M} = 0.094$.

\subsection{Testing the volume sampling}
\label{S:num-vol}

\begin{figure}
\centerline{\includegraphics[width=8.0cm]{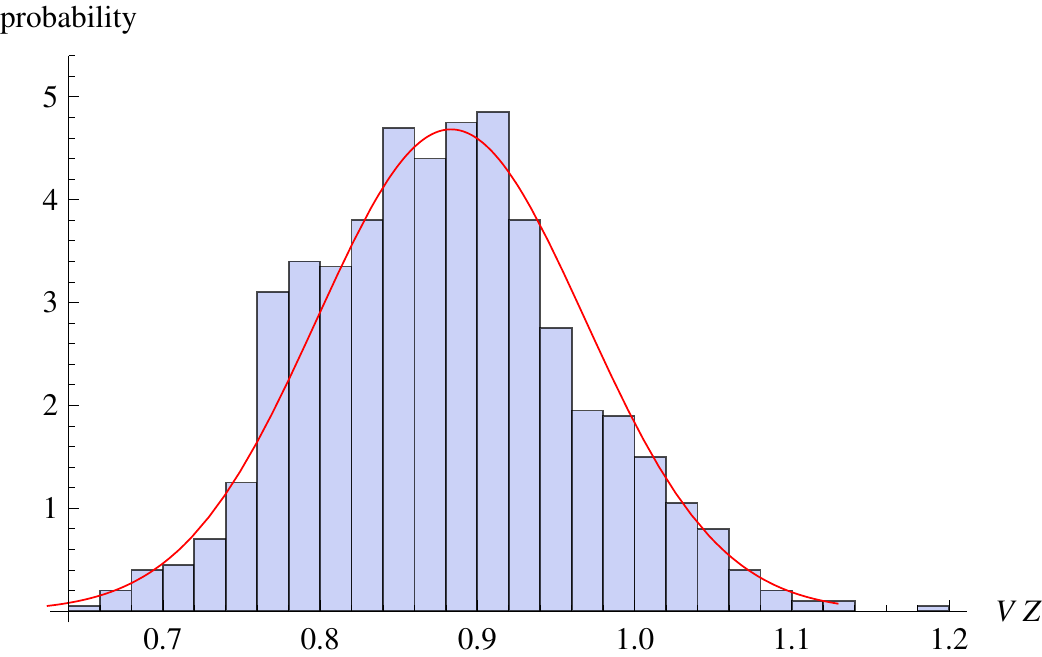}}
\caption{
The histogram shows the distribution of evidence values (specifically $V Z$) for 1000 realisations of the volume sampling, given a particular likelihood sampling.  The red curve shows a Gaussian distribution whose mean and varaiance are computed from \refeqs{Zavg}{Ksig}.  The mean and standard deviation from the simulations are $0.878\pm0.084$ (the mean differs from $V Z \approx 1$ for this particular likelihood sampling).  For comparison, the analytic average over volume realisations is 0.883, and Skilling's and my uncertainty estimators yield 0.083 and 0.085, respectively.
}\label{fig:volerr}
\end{figure}

I first generate a single realisation of the likelihood sampling and combine it with 1000 realisations of the volume sampling.  \reffig{volerr} shows a histogram of the $V Z$ values from these direct simulations.  The mean and standard deviation of the simulated values are $0.878\pm0.084$.  The mean differs from the theoretical value $V Z \approx 1$ by about $1.5\sigma$ for this particular realisation of the likelihood sampling.

From \refeq{Zavg} the predicted average over volume realisations is 0.883.  Skilling's estimator yields a statistical uncertainty of 0.083, while mine yields 0.085.  The predicted Gaussian distribution agrees well with the simulation results, indicating that $Z$ has a (nearly) Gaussian distribution when the uncertainties are small (qv.\ \refsecs{skilling}{moments}).  I conclude that the analytic expressions accurately describe the distribution of evidence values for many realisations of the volume sampling.  It is striking that the two uncertainty estimators yield very similar values despite having such different analytic forms.

\subsection{Testing the likelihood sampling}
\label{S:num-like}

\begin{figure}
\centerline{\includegraphics[width=8.0cm]{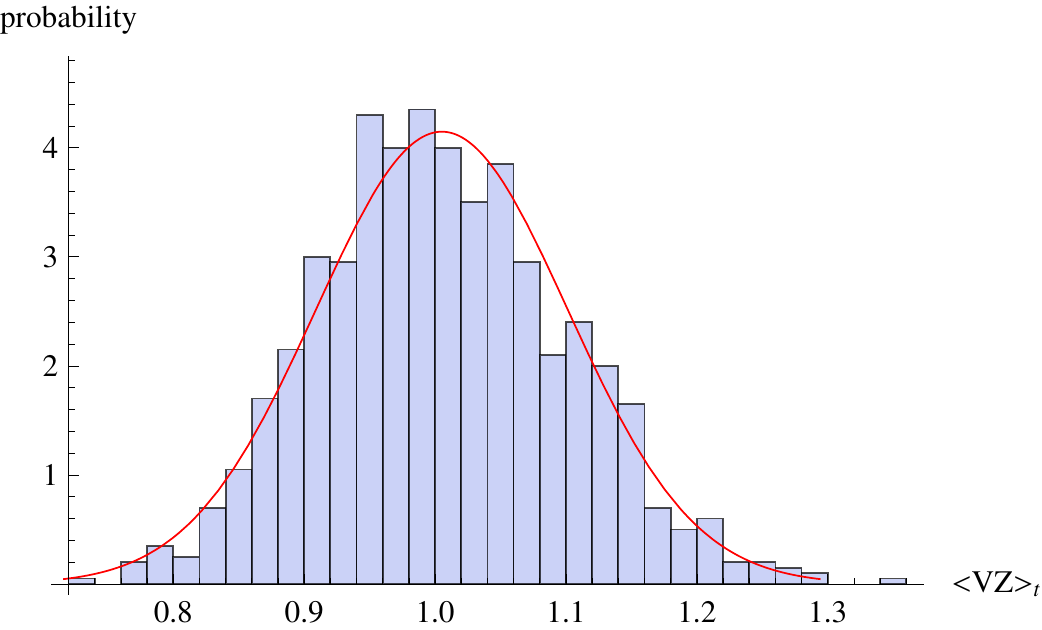}}
\caption{
The histogram shows the distribution of $\avg{VZ}_t$ for 1000 realisations of the likelihood sampling.  The notation $\avg{VZ}_t$ emphasises that I average over all volume samplings for each likelihood sampling.  The red curve shows a Gaussian distribution whose mean and variance are computed from \refeqs{Zavg}{Ksig}.  The mean over all likelihood samplings is 1.005; the empirical scatter in the histogram is 0.094, while the predicted value is 0.094 for Skilling's estimator and 0.096 for mine.
}\label{fig:histZavg}
\end{figure}

It is useful to see how the results vary with different realisations of the likelihood sampling.  I now generate 1000 random likelihood samplings; for each one I compute the mean evidence averaged over all volume samplings using \refeq{Zavg}.  \reffig{histZavg} shows a histogram of the values of $\avg{VZ}_t$ for the different likelihood realisations (I add the subscript $t$ to emphasise that the average is over volume samplings).  The mean and standard deviation for the histogram are $1.005\pm0.094$.  On average, nested sampling recovers the evidence very well.

Strictly speaking, both of the uncertainty estimators depend on the likelihood sampling, but the scatter across the likelihood realisations is $<9\%$ so any single case provides a useful value.  The average predicted uncertainty is 0.094 for Skilling's estimator, and 0.096 for mine.  Also, the predicted Gaussian distribution agrees well with the empirical histogram.  I conclude that the both analytic estimators characterise the statistical uncertainty in the evidence quite well.  It is not obvious at this point why the two uncertainty estimators yield such similar results.

\subsection{How many live points and steps?}
\label{S:num-steps}

Let us now consider how to choose the number of live points, $M$, and the number of nested sampling steps, $\Nnest$.  One general goal is to have nested sampling ``find'' all significant modes in the posterior.  The sampling procedure is basically guaranteed to find the peak for a unimodal distribution, but if the live points are too sparse they may miss some peaks (especially small ones) in a multi-model distribution.  In order to have a reasonable probability of getting at least one live point in each mode at the outset, \citet{Feroz1} suggest that the number of live points should exceed $V_{\rm prior}/V_{\rm min}$, where $V_{\rm prior}$ is the volume spanned by the priors while $V_{\rm min}$ is the volume of the smallest mode (which must be estimated since it cannot actually be known before the analysis is done).

\begin{figure}
\centerline{
  \includegraphics[width=8.0cm]{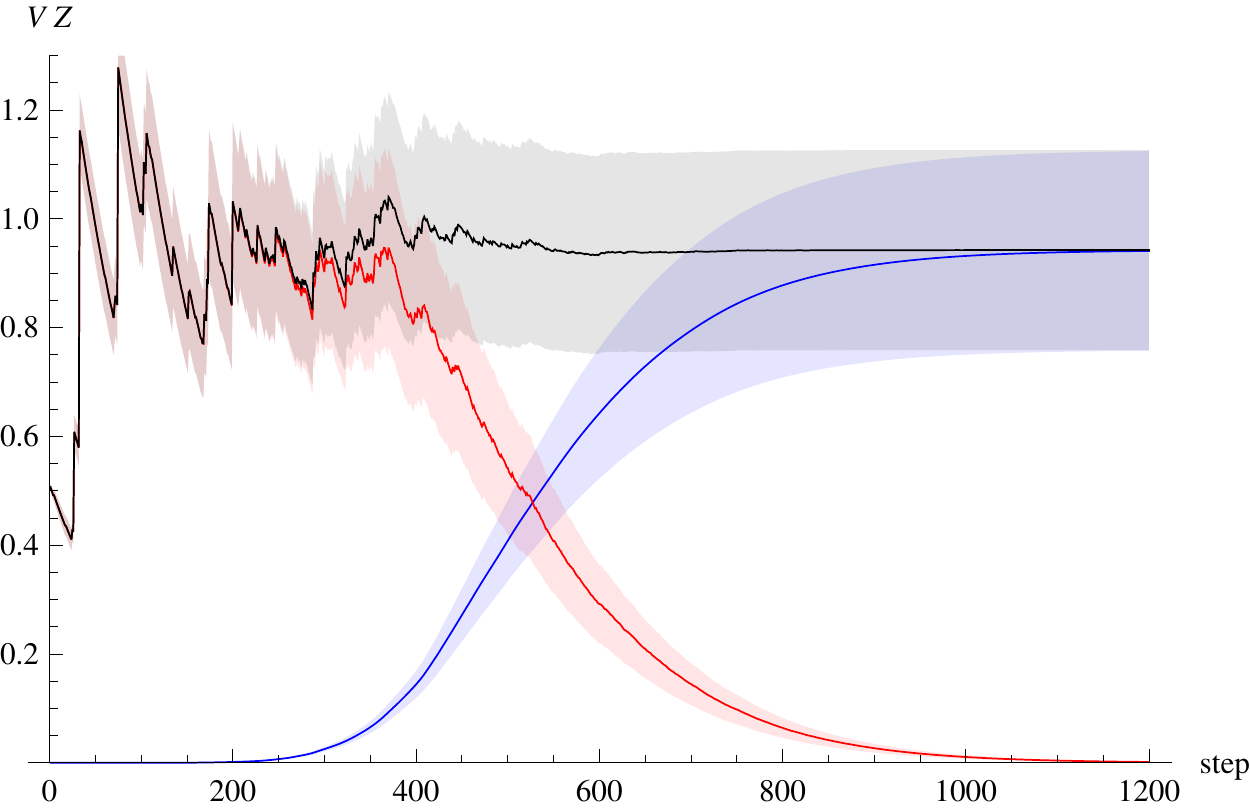}
  \hspace{0.2cm}
  \includegraphics[width=8.0cm]{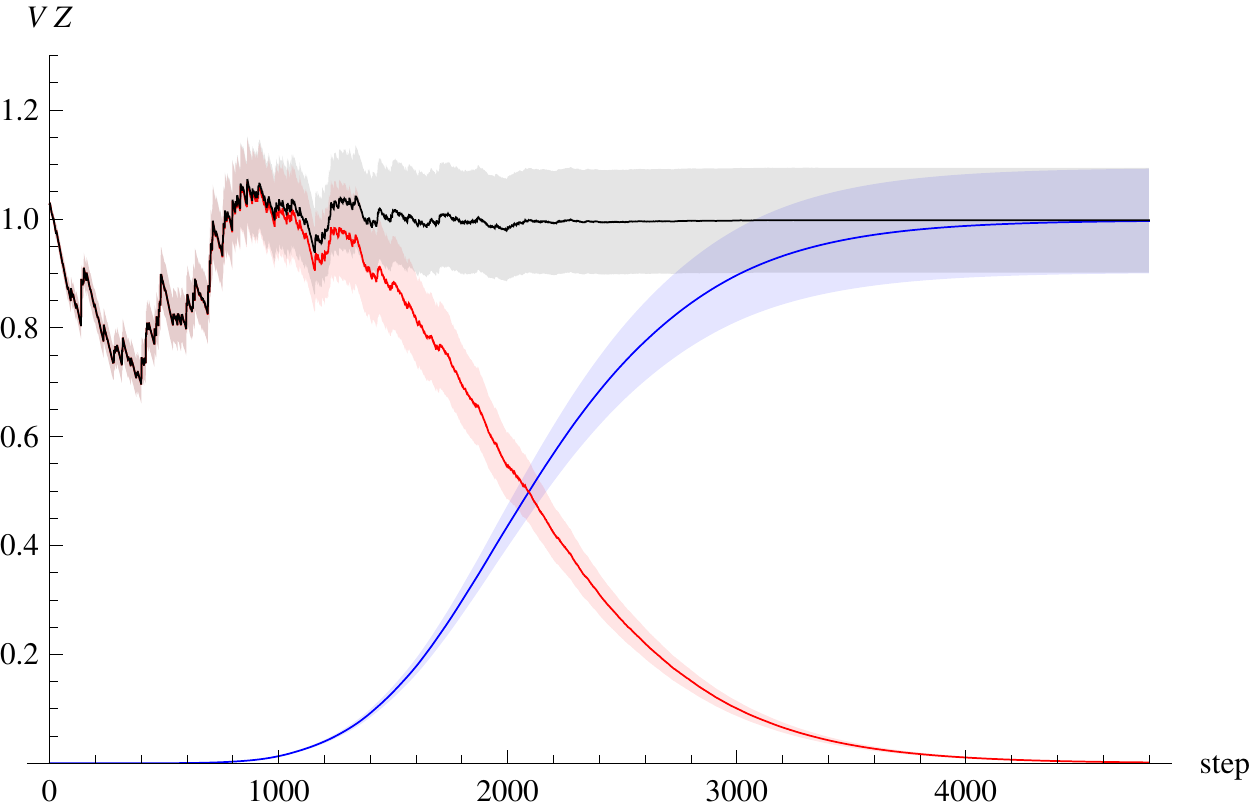}
}
\caption{
Development of the evidence as a function of step index.  The blue band shows the mean and $1\sigma$ errors for the ``nested evidence'' (eqs.~\ref{eq:Zavg} and \ref{eq:Ksig}), the red band shows the ``live evidence'' (eqs.~\ref{eq:avgZlive} and \ref{eq:sigZlive}), and the black curve shows the total (eqs.~\ref{eq:avgZtot} and \ref{eq:sigZtot}).  The number of live points is $M=100$ (left) and $M=400$ (right).  With more live points, it takes more steps to reach convergence, but the ultimate uncertainty is smaller.
}\label{fig:partevid}
\end{figure}

The second consideration relates to achieving a robust and precise estimate of the evidence.  \reffig{partevid} shows the development of the evidence as a function of the step index, for two choices of $M$.  After some number of steps the evidence and uncertainty saturate in the sense that taking additional steps does not significantly change the results.  For a heuristic understanding, note that as nested sampling homes in on a likelihood peak the likelihood values become constant ($L_i \to L_{\rm peak}$) while the volumes become progressively smaller ($X_i \to 0$).  For a rigorous proof of convergence, see \citet{Skilling09}.

The question arises of how to identify the point of diminishing returns.  One simple possibility \citep{Skilling04,Skilling06} is to compare the evidence accumulated through nested sampling ($Z_k$ from eq.~\ref{eq:partevid}) with the remaining evidence estimated from the live points ($\Zlive_k$ from eq.~\ref{eq:Zlive}).  \reffig{partevid} shows $\Zlive_k$ versus $k$ in red.  At early stages the curve shows a series of spikes: it rises sharply when a new live point is found that (temporarily) dominates the average over live points, then declines as the volume decreases.  The curve smoothes out as the live points come to have more similar likelihoods, and then decays as the likelihoods saturate while the remaining volume continues to decrease.  Roughly speaking, $\Zlive$ represents the evidence that has been ``missed'' by the nested sampling procedure, so we may want to continue the nested sampling until the ratio of live to nested evidence falls below some threshold: $\Zlive/Z < \epsilon$.

\begin{figure}
\centerline{\includegraphics[width=8.0cm]{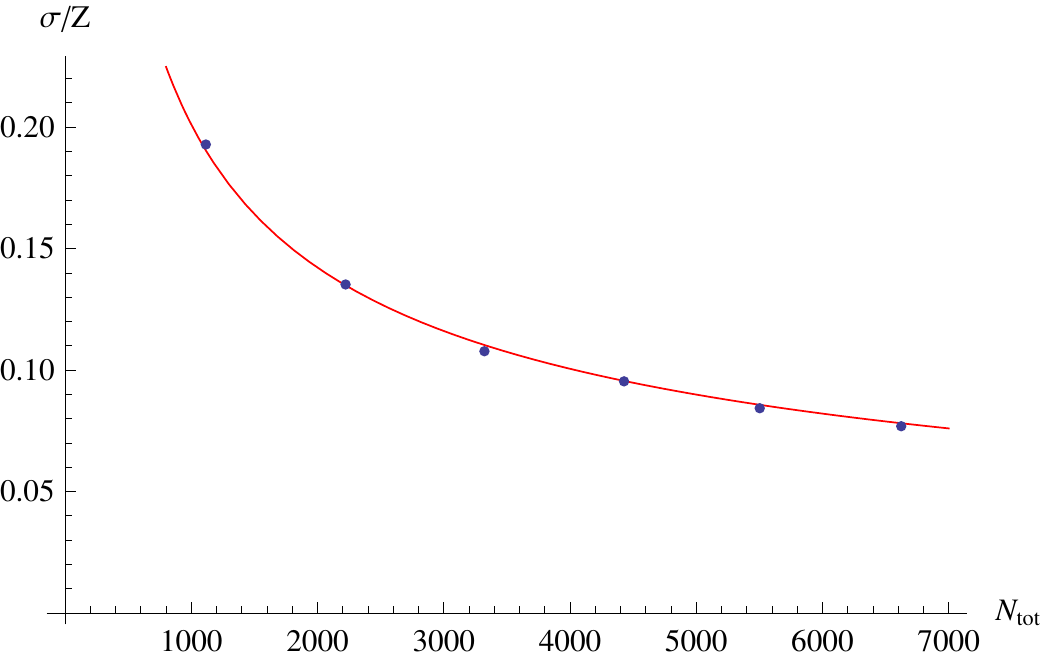}}
\caption{
The points show the fractional uncertainty in the evidence versus the total number of likelihood evaluations ($\Ntot = \Nnest+M$) for tests in which the number of live points is $M = 100, 200, 300, 400, 500, 600$ (left to right).  Here $\Nnest$ is determined for each $M$ using the stopping threshold $\epsilon = 0.01$.  The curve shows the scaling relation $\sigma_Z/Z \propto \Ntot^{-1/2}$.
}\label{fig:sig-vs-N}
\end{figure}

\reffig{partevid} illustrates that using more live points means it takes more steps to reach a given $\epsilon$ threshold, but the extra computational effort is rewarded with a smaller statistical uncertainty.  It is therefore interesting to compare the achieved uncertainty with the computational effort, which we may measure as the total number of likelihood samples ($\Ntot = \Nnest+M$).  \reffig{sig-vs-N} shows this comparison for different numbers of live points, given a fixed stopping threshold $\epsilon = 0.01$.  The fractional uncertainty clearly decreases with the total number of samples as $\sigma_Z/Z \propto \Ntot^{-1/2}$, just as expected for a statistical sampling procedure \citep{Skilling04,Skilling06,Chopin}.

In the examples presented here, I have used a low $\epsilon$ threshold to require that the live evidence be negligible at the end of the run.  \reffig{partevid} suggests, however, that $\epsilon$ could be set higher provided that the live evidence is accounted for properly (using the methods in \refsec{Zlive}).

The lessons here are familiar from previous work on nested sampling, but worth reiterating.  The ultimate statistical uncertainty depends mainly on the number of live points.  Once the nested sampling procedure has converged (as measured, for example, by the $\epsilon$ threshold), running more steps will not improve the results.  The way to reduce the uncertainties is to increase the number of live points.\footnote{It is not necessary to start from scratch in order to increase the number of live points.  \citet{Skilling06} explains that independent runs with $M_1$, $M_2$, $\ldots$ live points can be merged into a joint run that effectively has $M_1+M_2+\ldots$ live points.  The likelihood samplings are simply merged and sorted, while the volume sampling must be recomputed.}  That will increase the number of steps it takes to reach convergence, but will yield uncertainties that scale as $\sigma_Z \propto \Ntot^{-1/2}$.

\subsection{Log-normal test case}
\label{S:log}

Nowhere in the theoretical framework was it necessary to specify the form of the likelihood, so the Gaussian test case should be sufficient to validate the analytic results.  Nevertheless, it is useful to consider a different test to verify that the results are indeed robust.  I use a multivariate log-normal distribution because it is skewed and non-Gaussian but still analytically tractable.  Choosing appropriate scaled coordinates, we can write the likelihood in canonical form,
\begin{equation}
  \like(\theta) = \prod_{i=1}^{d} \frac{e^{-(\ln\theta_i)^2/2}}{(2\pi)^{1/2} \theta_i}
\end{equation}
where $d$ is the number of dimensions.  Let the prior volume be the cube with $0<\theta_i<s$, so $V = s^d$ and the evidence is
\begin{equation}
  Z = V^{-1} \prod_{i=1}^{d} \int_{0}^{s} \frac{e^{-(\ln\theta_i)^2/2}}{(2\pi)^{1/2} \theta_i}\ d\theta_i
\end{equation}
For a large prior box, $V Z \approx 1$ and the information gain is
\begin{equation}
  H = \frac{1}{V Z} \int \like \ln\frac{\like}{Z}\ d\theta
  \approx -\frac{d}{2}\left(1+\ln 2\pi\right) - \ln Z
\end{equation}
I again work in $d = 4$ dimensions, but now use a box with $s = 20$ to encompass the bulk of the likelihood.  Given the larger volume, I use $M = 600$ lives points and take $\Nnest = 9000$ steps.  For these parameter choices, the information gain is $H = 6.31$ and Skilling's estimator for the fractional uncertainty in the evidence has an analytic value of $\sqrt{H/M} = 0.103$.

I first consider a single likelihood sampling and examine the distribution of evidence values for 1000 volume realisations (qv.\ \refsec{num-vol}).  The empirical mean and standard deviation over the volume samplings are $0.939\pm0.097$.  The analytic mean is 0.944, while Skilling's and my estimators predict uncertainties of 0.097 and 0.098, respectively.  The histogram of $Z$ values (not shown) agrees well with a Gaussian distribution whose mean and variance are given by \refeqs{Zavg}{Ksig}.  (Note that $Z$ can have a nearly-Gaussian distribution even if the likelihood is non-Gaussian.)

I next consider 1000 likelihood samplings and examine the distribution of $\avg{VZ}_t$ values (qv.\ \refsec{num-like}).  The empirical mean and standard deviation are $0.997\pm0.106$.  Skilling's and my estimators predict uncertainties of 0.102 and 0.103, respectively.  The histogram of $\avg{VZ}_t$ again agrees well with the predicted Gaussian distribution.  I conclude that the analytic results are reliable even for a non-Gaussian likelihood distribution.

\section{Summary}
\label{S:summary}

I have derived simple analytic expressions for the mean and variance of the Bayesian evidence over all realisations of the volume sampling in nested sampling, and compared them with the uncertainty estimator introduced by \citet{Skilling04,Skilling06} from an information theoretic argument.  The two estimators have different forms as sums over the likelihood sampling, yet they yield very similar quantitative results.  At this point it is not clear whether the agreement reflects some general equivalence between the two estimators that is not yet apparent, or whether it somehow depends on statistical properties of the likelihood sampling $\{L_i\}$ that emerges from the nested sampling procedure.  The moments-based estimator currently has a more rigorous foundation than the information theoretic estimator, but both are useful and it will be interesting to see if they continue to give similar results as nested sampling is applied to a broader range of problems, and if any formal equivalence can be established.  Both estimators can be used to compute the statistical uncertainty in the evidence for any implementation that maintains the core prescription of nested sampling: each new point is drawn from the prior distribution in the region inside the current likelihood surface.  With these results, determining not only the mean evidence but also the uncertainty requires no additional computational effort (and no guesswork) beyond that needed to generate the likelihood sampling.

\section*{Acknowledgments}
\label{S:acknowledgments}

I thank Ross Fadely for valuable discussions about nested sampling, and the referee for suggesting the error analysis in \refsec{Zlive}.
This work received support from the US National Science Foundation through grant AST-0747311.


\bibliographystyle{mn2e}
\bibliography{keeton-nest-v2}

\end{document}